# Java Prolog Interface

Jose E. Zalacain Llanes[1]

## Abstract

There are many initiatives in presents-days for interaction between Java and Prolog programming languages. These initiatives allow combine two programming paradigms, Object Oriented Programming and Logic Programming. Every proposed interface has specifics features depending of the final use. The present paper introduces a new Java Prolog Interface to be use for Prolog persistence interacting from Java side and functional programming from Prolog side. To support this interaction, the most advanced solutions implements interlanguages data type mappings between Java objects and Prolog terms. Java Prolog Interface is a modern solution that take the best features from existing solutions and combine all in one. It's more flexible, adaptive and have an Application Provider Interface (API) easy to use. JPI implement the `javax.script` interface include in Java from version 1.6. The project like existing solutions have an implementation for the most popular open source Prolog Engines. Is hosted on GitHub[2] source code management at Prolobjectlink repository and deploy the resulting artifacts on Maven Central[3] repository. The project have a web page too hosted on GitHub[4].

## Introduction

There are many initiatives in presents-days for interaction between Java and Prolog programming languages. These initiatives allow combine two programming paradigms, Object Oriented Programming (OOP) and Logic Programming (LP). The currents solutions can be categorized in three types (Java-Based Prolog Engine implementations, specific Engine implementation and Multi-engine implementations). Java Prolog Interface (JPI) is categorized like Multi-engine implementation. Is influenced by Interprolog (Calejo, 2004), Prolog Development Tool Connector (PDT) (Rho, 2004) and Java Prolog Connectivity (JPC) (Sergio Castro, 2013), (Sergio Castro, 2014). Previous solutions only communicate native prolog engine implementations like SWI (Wielemaker, Schrijvers, & Triska, 2012), YAP (Victor Santos Costa, 2012) and XSB (Warren, 2012). JPI is an intent to connect all prolog implementations, Natives and Java-Based.

---

[1] jzalacain@nauta.cu

[2] https://github.com/prolobjectlink/prolobjectlink-jpi
[3] https://repo.maven.apache.org/maven2
[4] https://prolobjectlink.github.io/prolobjectlink-jpi

All Java Prolog connector contains in your design concepts like Engine Abstraction, Data Types and Interlanguage Type Converter (Mapping). The Engine Abstraction is the feature that allow interact with concrete engines in transparent way. It allows the coupling and decoupling different concrete Prolog engines implementations and the application code still remain the same. Only change the concrete Prolog Driver or Provider. The data types in Java-Prolog connector are an abstraction too, but for concrete data types. The main target for data types abstraction is similar to engine abstraction, code one. A special data type implementation is Reference data type (REF). This feature is essential for communicate Prolog to Java and hold Java references to prolog side. Interlanguage type converter is the feature that identify similar type and convert from one language to the other equivalent. It possibility formulate queries in one target language passing the queries to the other one and obtain the result in original language built-ins types. This feature is a modern approach described in JPC and CAPJA (L. Ostermayer, 2014).

JPI introduce new features doing a special difference over your influences. The Java Scripting Engine implementation. The Java Scripting Engine (Grogan, 2006), (Kashan, 2014), (Oracle, 2017) was introduced in Java Runtime Environment (JRE) in 1.6 version and release under `javax.script` package. The first intent to implement this interface was TuProlog (Denti, 2001) in 3.0.0 version. The Java Scripting Engine implementation is a possible use over one Java Specification API, but Java Prolog connector need more over Java Platform. Java Prolog Interface need be a specific API specification like (`javax.logic`) for all logic and functional language interaction. One specification like this allow more direct implementations reducing the impedance mismatch cost and self-platform distribution API. There are many project that use Java Prolog connectors and always a new interface is produced, JPL (Wielemaker & Angelopoulos, 2012), Interprolog, PDT Connector, JPC, CAPJA. A Java Prolog Specification will remove the repetitive interface creation.

## Architecture

JPI use a layered architecture pattern where every layer represents a component. The multi-engine Java Prolog connectors provide different levels of abstraction to simplify the implementations of common inter-operability task JPC. Java Prolog Connectors architectures describe three fundamentals layers, High-level API layer, Engine Adapter layer and Concrete Engine layer. High-level API layer define all services to be used by the users in the Java Prolog Application that is the final architecture layer on the architecture stack. High-level API provide the common implementation of Engine Abstraction, Data Type and Inter-Language conversion. The adapter layer adapts before mentioned features to communicate with the concrete Engine Layer, being the last responsible of execute the request services.

All existing Java Prolog Connectors implementation only bring support for Native Prolog Engines that have JVM bindings driver. JPI project is more inclusive and find connect all Prolog Engines Categories, Native and Java Based implementations. Some particular Java Based implementations in the future can be implement in strike forward mode the JPI interface. This particulars implementations reduce the impedance mismatch by remove the adapter layer. Therefore, JPI reference implementations will be faster than other that use adapter layer.

| User Application |||||
|---|---|---|---|---|
| Java Prolog Interface (JPI) |||||
| JPI-JPL | JPI-JTrolog | JPI-JLog | JPI-TuProlog | JPI- JPL7 |
| JPL | JTrolog | JLog | TuProlog | JPL7 |
| SWI | Java Virtual Machine (JVM) |||  SWI7 |
| Operating System |||||

Tab. 1: Component stack that represent the JPI architecture.

In JPI architecture stack in the bottom layer we have the Operating System. The Operating System can be Windows, Linux or Mac OS. Over Operating System, we have the native implementation of JVM and Prolog Engines like SWI, SWI7 and others. Over JVM and Prolog Engines we have Java Based Prolog Engines implementations and JVM bindings driver that share the runtime environment with JVM and native Prolog Engines. Over Java Based Prolog Engines implementations and JVM bindings drivers we have the JPI correspondent adapters. The adapters artifacts are the JPI implementations for each Prolog Engines. Over each adapter we have the JPI application provider interface and at the top stack we the final user application. The user application only interacts with the JPI providing single sourcing and transparency.

## Data Type Conversion

Many authors study the Java-Prolog inter language conversion JPC (Sergio Castro, 2013), (Sergio Castro, 2014), (Cimadamore & Viroli, 2007), (Cimadamore & Viroli, 2008), (Zalacain Llanes, 2017), CAPJA (L. Ostermayer, 2014). This work proof that is possible find a function $F$ over prolog data type set $T$ such that $F(t) = o$ and inverse function $F^{-1}$ over Java primitive data type set O such that $F^{-1}(o) = t$ where $t \in T$ and $o \in O$. The object term mapping can be resume in a single table.

| Logic Programing | Object Oriented |
|---|---|
| Nil | Null |
| True | True |
| Fail | False |
| Atom | String |
| Float | Float |
| Integer | Integer |
| Structure | Object |
| List | Array |

Tab. 2: Correspondence between LP data type and OO primitive data type.

JPC remark that it is not always possible find an inverse function to convert from Java primitive types to Prolog data type. This is because there are more primitive types in Java than Prolog. Byte, Short, Integer and Long produce a Prolog Integer data type but Prolog Integer only can produce Integer Java primitive type. The same case occurs with Character and String that only produce Prolog Atom, but Prolog Atom only produce Java String. To mitigate this problem is use the typed conversion technique that allow specify the result type. A special conversion case is multivalued conversion. Multivalued conversion allow convert List (Array) to Prolog List. JPI convert Prolog List to Java array of objects and the inverse case is supported too. Other solutions allow direct conversion to java.util.List but we consider that Java List it is not a primitive type. Another special case is Prolog Structure type that for this implementation level is itself represented like Java Object. There are other strategies at most advanced implementation level that map a Prolog Structure to some object instance of some related Class in correspondence to the most general predicate for the structure. Such is the case of CAPJA. Prolobjectlink Project have an equivalent implementation where this conversion is supported but it is not covered in this paper.

## Application Provider Interface

The JPI design use the common implementation technique for application provider's interfaces. JPI define an interface for every Prolog component and data type. Are derived from this interfaces abstract classes to join all common methods for concretes implementation classes. The concretes implementation classes define the particular way to implement some behavior. The table 3 and 4 show the correspondent interfaces and classes in JPI design.

| Interface | Description |
|---|---|
| **PrologAtom** | Represent the Prolog atom data type. |
| **PrologClause** | Prolog clause is composed by two prolog terms that define a prolog clause, the head and the body. |
| **PrologClauseBuilder** | Prolog clause builder to create prolog clauses. |
| **PrologClauses** | Clause family list that join all clauses with same functor/arity based indicator. |
| **PrologConsole** | Represent the prolog console of the system. |
| **PrologConverter**<T> | Converter for convert PrologTerm to the equivalent native T term representation. |
| **PrologDouble** | Prolog term that represent a double precision floating point number. |
| **PrologEngine** | A PrologEngine instance is used in order to interact with the concrete prolog engine. |
| **PrologFloat** | Prolog term that represent a single precision floating point number. |
| **PrologFormatter** | Class that define the string format for prolog logger output. have a format method that give some log record and return the string format for this record to be print in logger output. |
| **PrologIndicator** | Indicator to denote the signature for Prolog Terms using a functor/arity format. |
| **PrologInteger** | Prolog term that represent a integer number. |
| **PrologJavaConverter** | Converter for convert PrologTerm to the equivalent Java object taking like reference the following equivalence table. |
| **PrologList** | Represent prolog list compound term. |
| **PrologLogger** | Logger platform interface to log message at any level. |
| **PrologLong** | Prolog term that represent a long integer number. |
| **PrologNumber** | Represent all Prolog number data type. |
| **PrologOperator** | This class defines a Prolog operator. |
| **PrologOperatorSet** | A collection that contains no duplicate Prolog operators. |
| **PrologProvider** | Prolog Provider is the class to interact with all prolog components (data types, constants, logger, parser, converter and engine). |
| **PrologQuery** | Prolog query is the mechanism to query the prolog database loaded in prolog engine. |
| **PrologQueryBuilder** | Prolog query builder to create prolog queries. |
| **PrologReference** | Compound term that have like argument the object identification atom. |
| **PrologStructure** | Represent structured prolog compound term. |
| **PrologTerm** | Ancestor prolog data type. |
| **PrologVariable** | Prolog term that represent variable data type. |

Tab. 3: JPI interface summary.

| Class | Description |
|---|---|
| **AbstractClause** | Partial implementation of PrologClause interface. |
| **AbstractConsole** | Partial implementation of PrologConsole interface. |
| **AbstractConverter**<T> | Partial implementation of PrologConverter interface. |
| **AbstractEngine** | Partial implementation of PrologEngine. |
| **AbstractIndicator** | Partial implementation of PrologIndicator interface. |
| **AbstractIterator**<E> | Partial implementation of Iterator interface. |
| **AbstractJavaConverter** | Partial implementation of PrologJavaConverter interface. |
| **AbstractLogger** | Partial implementation of PrologLogger interface. |
| **AbstractOperator** | Partial implementation of PrologOperator. |
| **AbstractProvider** | Partial implementation of PrologProvider |
| **AbstractQuery** | Partial implementation of PrologQuery interface. |
| **AbstractReference** | Partial implementation of PrologReference interface. |
| **AbstractTerm** | Partial implementation of PrologTerm interface. |
| **ArrayIterator**<E> | Iterator implementation over array of elements. |
| **Licenses** | Class that contains some constants licenses names. |
| **Prolog** | Bootstrap platform class. |
| **PrologScriptEngineFactory** | Partial implementation of ScriptEngineFactory |
| **PrologTermType** | Contains all PrologTerm types constants |

Tab. 4: JPI class summary.

JPI describe a group situation that produce errors. For this situations the prolog interface has several error classes can be raised when some error occurs. The table 5 show the correspondent interfaces and classes in JPI design.

| Error | Description |
|---|---|
| **ArityError** | Runtime error raised when occurs one call to get arity method over a term that no have arity property. |
| **CompoundExpectedError** | Runtime error raised when occurs one call to some method over no compound term like get arguments or get argument at some position. all atomics term no have arguments and optionally over related invocations of the mentioned methods this runtime error take place. |
| **FunctorError** | Runtime error raised when occurs one call to get functor method over a term that no have functor property. |
| **IndicatorError** | Runtime error raised when occurs one call to get indicator method over a term that no have indicator property. |
| **ListExpectedError** | Runtime error raised when the expected term is a Prolog list. |
| **PrologError** | Common runtime error that can be used for any Prolog error notification. |
| **StructureExpectedError** | Runtime error raised when the expected term is a Prolog structure. |
| **SyntaxError** | Runtime error raised when occurs one syntax error. |

| Modifier and Type | Method and Description |
|---|---|
| **UnknownTermError** | Runtime error raised when **PrologConverter** don't have an equivalent term for some passed object. |

Tab. 5: JPI error summary.

# Prolog Converter

PrologConverter is a converter for convert PrologTerm to the equivalent native term representation. Contains several methods can be used in different conversion situation. The interface methods describe two method categories. The methods that convert from PrologTerm returning the equivalent native term representation and The methods that convert to PrologTerm using the equivalent native term representation. Some PrologConverter class conversion methods can receive as a second parameter the expected type of the converted object. This method category describes the Typed Conversion technique. A special method category is Multi-Valued Conversion. This category of converters also provides conversions for multi-valued data types such as arrays. The Typed Conversion and Multi-Valued Conversion are conversion techniques mentioned in (Sergio Castro, 2014).

| Modifier and Type | Method and Description |
|---|---|
| **PrologProvider** | **createProvider**()<br>Create a Prolog provider instance. |
| **T** | **fromTerm**(**PrologTerm** term)<br>Create a native term representation from given Prolog term. |
| <K> K | **fromTerm**(**PrologTerm** term, **Class**<K> to)<br>Create a native rule representation term from given head and body and cast this native term to some specific given class. |
| **T** | **fromTerm**(**PrologTerm** head, **PrologTerm**[] body)<br>Create a native rule representation term from given head and body. |
| <K> K | **fromTerm**(**PrologTerm** head, **PrologTerm**[] body, **Class**<K> to)<br>Create a native rule representation term from given head and body and cast this native term to some specific given class. |
| **T**[] | **fromTermArray**(**PrologTerm**[] terms)<br>Create a native term array representation from given Prolog term array. |
| <K> K[] | **fromTermArray**(**PrologTerm**[] terms, **Class**<K[]> to) |

|  | Create a native term array representation from given Prolog term array and cast this native term array to some specific given array class. |
|---|---|
| `Class<T>` | `getGenericClass()`<br>Get the generic class for the current Prolog converter at runtime. |
| `<K extends PrologTerm>`<br>`K` | `toTerm(Object o, Class<K> from)`<br>Create an equivalent Prolog term using the given native term representation and cast this Prolog term to some specific given class. |
| `PrologTerm` | `toTerm(T prologTerm)`<br>Create an equivalent Prolog term using the given native term representation. |
| `<K extends PrologTerm>`<br>`K[]` | `toTermArray(Object[] objects, Class<K[]> from)`<br>Create an equivalent Prolog terms array using the given native terms array representation and cast this Prolog term array to some specific array component class. |
| `PrologTerm[]` | `toTermArray(T[] terms)`<br>Create an equivalent Prolog terms array using the given native terms array representation. |
| `Map<String,PrologTerm>` | `toTermMap(Map<String,T> map)`<br>Create an equivalent Prolog terms map using the given native terms map representation. |
| `<K extends PrologTerm,V>`<br>`Map<String,PrologTerm>` | `toTermMap(Map<String,V> map, Class<K> from)`<br>Create an equivalent Prolog terms map using the given native terms map representation and cast every Prolog term to some specific given class. |
| `Map<String,PrologTerm>[]` | `toTermMapArray(Map<String,T>[] map)`<br>Create an equivalent Prolog terms map array using the given native terms map array representation. |
| `<K extends PrologTerm,V>`<br>`Map<String,PrologTerm>[]` | `toTermMapArray(Map<String,V>[] map, Class<K> from)`<br>Create an equivalent Prolog terms map array using the given native terms map array representation and cast every Prolog term to some specific given class. |
| `<K extends PrologTerm>`<br>`K[][]` | `toTermMatrix(Object[][] objects, Class<K[][]> from)`<br>Create an equivalent Prolog terms matrix using the given native terms matrix representation and cast every Prolog terms matrix to some specific matrix component class. |

| `PrologTerm[][]` | `toTermMatrix(T[][] terms)`<br>Create an equivalent Prolog terms matrix using the given native terms matrix representation. |
|---|---|

Tab. 6. Prolog Java Converter interface methods.

## Prolog Java Converter

PrologJavaConverter is the converter for convert PrologTerm to the equivalent Java object taking like reference the following equivalence table. The classes that implement this interfaces are responsible of the Primitive Conversions (Sergio Castro, 2014).

| Java Object | Prolog Term |
|---|---|
| `null` | `PrologProvider.prologNil()` |
| `String` | `PrologAtom` |
| `Boolean.FALSE` | `PrologProvider.prologFalse()` |
| `Boolean.TRUE` | `PrologProvider.prologTrue()` |
| `Integer` | `PrologInteger` |
| `Float` | `PrologFloat` |
| `Double` | `PrologDouble` |
| `Long` | `PrologLong` |
| `Object[]` | `PrologList` |

Tab. 7. Prolog Java Converter equivalence table.

There are special cases of Java object that can be converted to Prolog equivalent but the inverse case it's not possible. They are Byte, Character, Short that can be converted to PrologInteger using your numeric value. The main problems are that after PrologInteger conversion this value will be converted in Integer.

| Return Type | Method and Description |
|---|---|
| `boolean` | `containQuotes(String functor)`<br>Check if the current functor have quotes at the start and end of the given functor. |
| `String` | `removeQuotes(String functor)`<br>Remove functor quotes if they are present. |
| `Object` | `toObject(PrologTerm term)`<br>Create a Java object from given Prolog term. |
| `List<Object>` | `toObjectList(PrologTerm[] terms)`<br>Create a Java objects list from given Prolog term array. |
| `List<List<Object>>` | `toObjectLists(PrologTerm[][] terms)`<br>Create an equivalent list of objects lists using the given Prolog terms matrix. |
| `Map<String,Object>` | `toObjectMap(Map<String,PrologTerm> map)`<br>Create an equivalent Java object map using the given Prolog terms map. |
| `List<Map<String,Object>>` | `toObjectMaps(Map<String,PrologTerm>[] maps)` |

|              | Create an equivalent Java object map list using the given Prolog terms map array. |
|--------------|-----------------------------------------------------------------------------------|
| `Object[]`   | `toObjectsArray(PrologTerm[] terms)` Create a Java objects array from given Prolog term array. |
| `PrologTerm` | `toTerm(Object object)` Create an equivalent Prolog term using the given Java object. |
| `PrologTerm[]` | `toTermsArray(Object[] objects)` Create an equivalent Prolog terms array using the given Java objects array. |

Tab. 8. Prolog Java Converter interface methods.

# Prolog Provider

Prolog Provider is the mechanism to interact with all Prolog components. Provider classes implementations allow create Prolog Terms, Prolog Engine, Java Prolog Converter, Prolog Parsers and system logger. Using `org.prolobjectlink.prolog.Prolog` bootstrap class the Prolog Providers are created specifying the provider class in `getProvider(Class<?>)` method. This is the workflow start for JPI. When the Prolog Provider is created the next workflow step is the Prolog Terms creation using Java primitive types or using string with Prolog syntax. Provider allow create/parsing all Prolog Terms (Atoms, Numbers, Variables and Compounds). After term creation/parsing the next step is create an engine instance with `newEngine()` method. Using previous term creation and engine instance Prolog Queries can be formulated. This is possible because the engine class have multiples queries creation methods like a query factory. After query creation the Query interface present many methods to retrieve the query results. The result methods are based on result quantities, result terms, result object types, etc… This is the final step in the workflow. In the table 10 is resumed all Prolog Provider Interface methods.

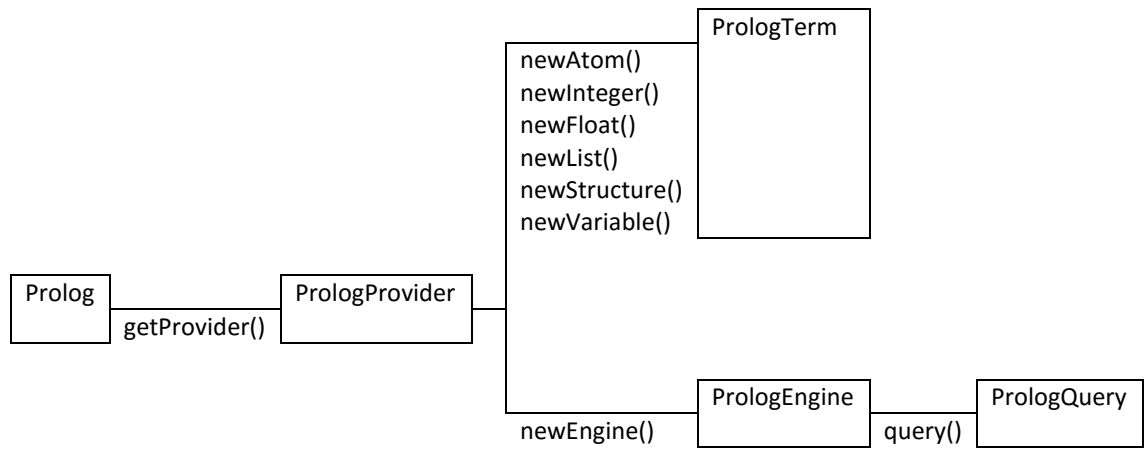

Tab. 9. Prolog workflow

| Return Type | Method and Description |
|---|---|
| `<K> K` | **fromTerm**(`PrologTerm` term, `Class`<K> to)<br><br>Create a native rule representation term from given head and body and cast this native term to some specific given class. |
| `<K> K` | **fromTerm**(`PrologTerm` head, `PrologTerm`[] body, `Class`<K> to)<br><br>Create a native rule representation term from given head and body and cast this native term to some specific given class. |
| `<K> K[]` | **fromTermArray**(`PrologTerm`[] terms, `Class`<K[]> to)<br><br>Create a native term array representation from given Prolog term array and cast this native term array to some specific given array class. |
| `<K> PrologConverter`<K> | **getConverter**()<br><br>Get a prolog converter instance to map the abstract prolog data types to under-laying prolog implementations data types. |
| `PrologJavaConverter` | **getJavaConverter**()<br><br>Get a Java to Prolog converter instance to map the abstract prolog data types to Java types. |
| `PrologLogger` | **getLogger**()<br><br>Get the prolog system logger instance to report any errors or exceptions |
| `String` | **getName**()<br><br>Name of the wrapped engine. |
| `PrologParser` | **getParser**()<br><br>Get a prolog parser instance to parser the strings with prolog syntax. |
| `String` | **getVersion**()<br><br>Version of the wrapped engine. |
| `boolean` | **isCompliant**()<br><br>True if wrapped engine implement ISO Prolog and false in other case |
| `PrologAtom` | **newAtom**(`String` functor) |

| | Create a prolog atom term setting like atom value the given string. |
|---|---|
| **PrologDouble** | **newDouble**() <br><br> Create a prolog double number instance with 0.0 value. |
| **PrologDouble** | **newDouble**(**Number** value) <br><br> Create a prolog double number instance with the given value. |
| **PrologEngine** | **newEngine**() <br><br> Create a new prolog engine instance ready to be operate. |
| **PrologEngine** | **newEngine**(**String** file) <br><br> Create a new prolog engine instance ready to be operate. |
| **PrologFloat** | **newFloat**() <br><br> Create a prolog float number instance with 0.0 value. |
| **PrologFloat** | **newFloat**(**Number** value) <br><br> Create a prolog float number with the given value. |
| **PrologInteger** | **newInteger**() <br><br> Create a prolog integer number instance with 0 value. |
| **PrologInteger** | **newInteger**(**Number** value) <br><br> Create a prolog integer number instance with the given value. |
| **PrologList** | **newList**() <br><br> Create an empty prolog list term. |
| **PrologList** | **newList**(**Object** head) <br><br> Create a prolog list with one object item. |
| **PrologList** | **newList**(**Object**[] arguments) <br><br> Create a prolog list from java objects arguments array and the tail item is an empty list. |
| **PrologList** | **newList**(**Object**[] arguments, **Object** tail) <br><br> Create a prolog list from java objects arguments array and the tail item is the given java object. |

| `PrologList` | `newList(Object head, Object tail)`<br><br>Create a prolog list with two java objects items [head \| tail]. |
|---|---|
| `PrologList` | `newList(PrologTerm head)`<br><br>Create a prolog list with one term item. |
| `PrologList` | `newList(PrologTerm[] arguments)`<br><br>Create a prolog list from prolog terms arguments array and the tail item is an empty list. |
| `PrologList` | `newList(PrologTerm[] arguments, PrologTerm tail)`<br><br>Create a prolog list from prolog terms arguments array and the tail item is the given prolog term. |
| `PrologList` | `newList(PrologTerm head, PrologTerm tail)`<br><br>Create a prolog list with two terms items [head \| tail]. |
| `PrologLong` | `newLong()`<br><br>Create a prolog long number instance with 0 value. |
| `PrologLong` | `newLong(Number value)`<br><br>Create a prolog long number instance with the given value. |
| `PrologTerm` | `newReference(Object object)`<br><br>Create a prolog object reference term that hold the given object. |
| `PrologTerm` | `newStructure(Object left, String operator, Object right)`<br><br>Create a prolog structure that represent an expression defined by your left and right operands separated by infix operator. |
| `PrologTerm` | `newStructure(PrologTerm left, String operator, PrologTerm right)`<br><br>Create a prolog structure that represent an expression defined by your left and right operands separated by infix operator. |
| `PrologTerm` | `newStructure(String functor, Object... arguments)` |

| | Create a prolog structure with the functor (structure name) and java objects arguments array. |
|---|---|
| `PrologStructure` | `newStructure(String functor, PrologTerm... arguments)` |
| | Create a prolog structure with the functor (structure name) and prolog terms arguments array. |
| `PrologVariable` | `newVariable(int position)` |
| | Create an anonymous variable instance with associated index. |
| `PrologVariable` | `newVariable(String name, int position)` |
| | Create an named variable instance with associated index. |
| `PrologClause` | `parseClause(String clause)` |
| | Parse the string with Prolog syntax and create an equivalent Prolog clause instance. |
| `PrologList` | `parseList(String stringList)` |
| | Parse the string with Prolog syntax and create an equivalent Prolog list term instance. |
| `Set<PrologClause>` | `parseProgram(File in)` |
| | Parse the Prolog text contained at specific file and return a Prolog clause set found in the file. |
| `Set<PrologClause>` | `parseProgram(String file)` |
| | Parse the Prolog text contained at specific file path and return a Prolog clause set found in the file. |
| `PrologStructure` | `parseStructure(String stringStructure)` |
| | Parse the string with Prolog syntax and create an equivalent Prolog structure term instance. |
| `PrologTerm` | `parseTerm(String term)` |
| | Parse the string with Prolog syntax and create an equivalent Prolog term instance. |
| `PrologTerm[]` | `parseTerms(String stringTerms)` |
| | Parse the comma separate terms in the given string with prolog syntax and return an array of terms formed by the comma separate terms. |
| `PrologTerm` | `prologCut()`<br><br>Get the prolog term that represent the prolog cut built-in. |
| `PrologTerm` | `prologEmpty()`<br><br>Get the prolog empty list term. |
| `PrologTerm` | `prologFail()`<br><br>Get the prolog fail term that represent fail built-in. |
| `PrologTerm` | `prologFalse()`<br><br>Get the prolog false term that represent false built-in. |
| `PrologTerm` | `prologInclude(String file)`<br><br>Get the prolog term representing the directive use by under-laying prolog implementation for file inclusion. |
| `PrologTerm` | `prologNil()`<br><br>Get the prolog nil term representing the null data type for prolog data type system. |
| `PrologTerm` | `prologTrue()`<br><br>Get the prolog true term that represent true built-in. |
| `<K extends PrologTerm> K` | `toTerm(Object o, Class<K> from)`<br><br>Create an equivalent Prolog term using the given native term representation and cast this Prolog term to some specific given class. |
| `<K extends PrologTerm> K[]` | `toTermArray(Object[] objects, Class<K[]> from)`<br><br>Create an equivalent Prolog terms array using the given native terms array representation and cast this Prolog term array to some specific array component class. |
| `<K extends PrologTerm,V> Map<String,PrologTerm>` | `toTermMap(Map<String,V> map, Class<K> from)`<br><br>Create an equivalent Prolog terms map using the given native terms map representation and cast every Prolog term to some specific given class. |

| `<K extends PrologTerm,V>`<br>`Map<String,PrologTerm>[]` | `toTermMapArray(Map<String,V>[] map, Class<K> from)`<br><br>Create an equivalent Prolog terms map array using the given native terms map array representation and cast every Prolog term to some specific given class. |
|---|---|
| `<K extends PrologTerm>`<br>`K[][]` | `toTermMatrix(Object[][] objects, Class<K[][]> from)`<br><br>Create an equivalent Prolog terms matrix using the given native terms matrix representation and cast every Prolog terms matrix to some specific matrix component class. |

Tab. 10. Prolog Provider Interface Methods

JPI reduce the code lines to interact with Prolog engines and language. Is very simple and easy to use this is showed in Hello World program.

```
public class Main {
	public static void main(String[] args) {
		PrologProvider provider = Prolog.getProvider(SwiProlog.class);
		PrologEngine engine = provider.newEngine();
		engine.asserta("sample('hello wolrd')");
		PrologQuery query = engine.query("sample(X)");
		System.out.println(query.one());
	}
}
```

# Prolog Terms

All Java Prolog connector libraries provide data type abstraction. Prolog data type abstraction have like ancestor the Term class. Prolog term is coding like abstract class and other Prolog terms are derived classes. In PrologTerm is defined the common term operation for all term hierarchy (functor, arity, compare, unify, arguments). The derived classes implement the correct behavior for each before mentioned operations. All Prolog data types PrologAtom, PrologNumber, PrologList, PrologStructure and PrologVariable are derived from this class. All before mentioned classes extends from this class the commons responsibilities. PrologTerm extends from Comparable interface to compare the current term with another term based on Standard Order[5]. The data type hierarchy for JPI is show in the figure 4 and in the table 5 is resumed PrologTerm methods.

PrologAtom represent the Prolog atom data type. Prolog atoms are can be of two kinds simple or complex. Simple atoms are defined like a single alpha numeric word that begin like initial lower case character. The complex atom is defining like any character sequence that begin and end with simple quotes. The string passed to build a simple atom should be match with [a-z] [A-Za-z0-9_] * regular expression. If the string passed to build an atom don't match with the before mentioned regular expression the atom constructor can be capable of create a complex atom automatically. For complex atom the string value can have the quotes or just can be absent. The printed string representation of the complex atom implementation set the quotes if they are needed.

---
[5] Variables < Atoms < Numbers < Compounds. The same type term is compared by value or alphabetic order.

```
PrologTerm pam = provider.newAtom("pam");
PrologTerm bob = provider.newAtom("bob");
```

PrologDouble represent a double precision floating point number. Extends from PrologNumber who contains an immutable Double instance. The Prolog Provider is the mechanism to create a new Prolog double invoking `PrologProvider.newDouble(Number)`. PrologFloat represent a single precision floating point number. Extends from PrologNumber who contains an immutable Float instance. The Prolog Provider is the mechanism to create a new Prolog float invoking `PrologProvider.newFloat(Number)`. PrologInteger represent an integer number. Extends from PrologNumber who contains an immutable Integer instance. The Prolog Provider is the mechanism to create a new Prolog integer invoking `PrologProvider.newInteger(Number)`. Prolog term that represent a long integer number. Extends from PrologNumber who contains an immutable Long instance. The Prolog Provider is the mechanism to create a new Prolog long integer invoking `PrologProvider.newLong(Number)`.

```
PrologTerm pi = provider.newDouble(Math.PI);
PrologTerm euler = provider.newFloat(Math.E);
PrologTerm i = provider.newInteger(10);
PrologTerm l = provider.newLong(10);
```

PrologVariable is created using `PrologProvider.newVariable(int)` for anonymous variables and `PrologProvider.newVariable(String, int)` for named variables. The Prolog variables can be used and reused because they remain in java heap. You can instantiate a prolog variable and used it any times in the same clause because refer to same variable every time. The integer parameter represents the declaration variable order in the Prolog clause starting with zero.

```
 PrologTerm x = provider.newVariable("X", 0);
 PrologTerm y = provider.newVariable("Y", 1);
 PrologTerm z = provider.newVariable("Z", 2);
 engine.assertz(
              provider.newStructure(grandparent, x, z),
               provider.newStructure(parent, x, y),
                provider.newStructure(parent, y, z)
                   );
```

PrologReference term is inspired on JPL JRef. This term is like a structure compound term that have like argument the object identification atom. The functor is the @ character and the arity is 1. An example of this prolog term is e.g. @(J#00000000000000425). To access to the referenced object, is necessary use `PrologTerm.getObject()`.

PrologList are a special compound term that have like functor a dot (.) and arity equals 2. Prolog list are recursively defined. The first item in the list is referred like list head and the second item list tail. The list tail can be another list that contains head and tail. A special list case is the empty list denoted by no items brackets ([]). The arity for this empty list is zero. The Prolog Provider is the mechanism to create a new PrologList is invoking `PrologProvider.newList()` for empty list or `PrologProvider.newList(PrologTerm)` for one item list or `PrologProvider.newList(PrologTerm[])` for many items.

```
 PrologTerm empty = provider.newList();
```

```
PrologTerm one = provider.newInteger(1);
PrologTerm two = provider.newInteger(2);
PrologTerm three = provider.newInteger(3);
PrologTerm list = provider.newList(new PrologTerm[] { one, two, three});
for (PrologTerm prologTerm : list) {
	System.out.println(prologTerm);
}
```

PrologList implement Iterable interface to be used in for each sentence iterating over every element present in the list.

```
Iterator<PrologTerm> i = list.iterator();
while (i.hasNext()) {
	PrologTerm prologTerm = i.next();
	System.out.println(prologTerm);
}

for (Iterator<PrologTerm> i = list.iterator(); i.hasNext();) {
	PrologTerm prologTerm = i.next();
	System.out.println(prologTerm);
}
```

Prolog structures consist in a relation the functor (structure name) and arguments enclosed between parenthesis. The Prolog Provider is the mechanism to create a new Prolog structures invoking `PrologProvider.newStructure(String, PrologTerm...)`. Two structures are equals if and only if are structure and have equals functor and arguments. Structures terms unify only with same functor and arguments structures, with free variable or with with structures where your arguments unify if they have the same functor and arity. Structures have a special property named arity that means the number of arguments present in the structure. There are two special structures term. They are expressions (Two arguments structure term with operator functor) and atoms (functor with zero arguments). For the first special case must be used `PrologProvider.newStructure(PrologTerm, String, PrologTerm)` specifying operands like arguments and operator like functor.

```
PrologTerm pam = provider.newAtom("pam");
PrologTerm bob = provider.newAtom("bob");
PrologTerm parent = provider.newStructure("parent", pam, bob);
```

| Term Super-interfaces | Ancestor Term | Term Sub-interfaces | Number Sub-interfaces |
|---|---|---|---|
| Comparable<PrologTerm> | PrologTerm | PrologAtom<br>PrologVariable<br>PrologReference<br>PrologList<br>PrologStructure<br>PrologNumber | |
| | | | PrologDouble<br>PrologFloat |

|  |  |  | PrologInteger PrologLong |
|---|---|---|---|

Tab. 11. Prolog Terms Taxonomy.

Other connectors have different type organization. For example, JPL and JPC have a remarkable difference in Lists and Structures creation. Both terms belong to Compound class and the creation of this terms specify the compound term functor. JPI have one class for each term. Coding using JPI is a strong typing way to interact with Prolog Terms.

| Return Type | Method and Description |
|---|---|
| **PrologTerm** | **getArgument**(int index)<br>Term located at some given index position in the current term arguments if current term is a compound term. |
| **PrologTerm**[] | **getArguments**()<br>Term arguments if the current term is a compound term. |
| int | **getArity**()<br>Term arity. |
| **String** | **getFunctor**()<br>Term functor. The functor of a term is a name for compound terms. |
| **String** | **getIndicator**()<br>Gets the term indicator represented by one string with the format functor/arity. |
| **Object** | **getObject**()<br>For references terms return the referenced object. |
| **PrologProvider** | **getProvider**()<br>Prolog provider associated to the current term. |
| **PrologTerm** | **getTerm**()<br>Return current term instance if current term is not a variable or is a free variable term. |
| int | **getType**()<br>Get the term type. |
| boolean | **hasIndicator**(**String** functor, int arity)<br>True if term has an indicator with the format functor/arity that match with the given functor and arity. |
| boolean | **isAtom**()<br>True if this term is an atom |
| boolean | **isAtomic**()<br>True if this Term is an atomic term, false in other case |
| boolean | **isCompound**() |
| | True if this Term is a compound term, false in other case |
| boolean | **isDouble**() <br> True if this Term is a double precision floating point number, false in other case |
| boolean | **isEmptyList**() <br> True if this Term is an empty list term ([]), false in other case |
| boolean | **isEvaluable**() <br> Check if the current term is a compound term and have like functor an operator. |
| boolean | **isFalseType**() <br> Check if the current term is a reference term and the referenced object is an instance of java false value. |
| boolean | **isFloat**() <br> True if this Term is a single precision floating point number, false in other case |
| boolean | **isInteger**() <br> True if this Term is an integer number, false in other case |
| boolean | **isList**() <br> True if this Term is a list, false in other case |
| boolean | **isLong**() <br> True if this Term is a large integer number, false in other case |
| boolean | **isNil**() <br> True if this Term is a nil term (null term for prolog), false in other case |
| boolean | **isNullType**() <br> Check if the current term is a reference term and the referenced object is a java null value. |
| boolean | **isNumber**() <br> True if this term is an number |
| boolean | **isObjectType**() <br> Check if the current term is a reference term for some java object instance. |
| boolean | **isReference**() <br> Check if the current term is a reference term for some java object instance or is a reference term and the referenced object is a java null value. |
| boolean | **isStructure**() <br> True if this Term is a structured term, false in other case |
| boolean | **isTrueType**() |

| | Check if the current term is a reference term and the referenced object is an instance of java true value. |
|---|---|
| boolean | **isVariable**()<br>True if this Term is a variable, false in other case |
| boolean | **isVoidType**()<br>Check if the current term is a reference term for java void type. |
| boolean | **unify**(**PrologTerm** term)<br>Check that the current term unify with the given term. |

Tab. 12. Prolog Term Interface Methods.

## Prolog Operator

Prolog operators are composed by a string operator name, string operator specifier or type and an operator priority. Extends from Comparable to compare with others operators instance over priority property.

| Return Type | Method and Description |
|---|---|
| **String** | **getOperator**()<br>String symbol that represent the Prolog operator. |
| int | **getPriority**()<br>Integer number between 0 and 1200 that represent the operator priority. |
| **String** | **getSpecifier**()<br>String symbol that specify the associativity and position of the Prolog operator. |

Tab. 13. Prolog Operator Interface Methods.

## Prolog Indicator

Indicator to denote the signature for Prolog Terms using a functor/arity format. More formally the indicator is formed by the concatenation of the term functor and term arity separated by slash.

| Return Type | Method and Description |
|---|---|
| int | **getArity**()<br>Indicator arity that is the argument number for compound terms. |
| **String** | **getFunctor**()<br>Indicator functor that is the name for compound terms. |
| **String** | **getIndicator**() |

| | Gets the term indicator represented by one string with the format functor/arity. |

Tab. 14. Prolog Operator Interface Methods.

# Prolog Engine

Prolog Engine provide a general propose application interface to interact with Prolog Programing Language. Is a convenient abstraction for interacting with Prolog Virtual Machine from Java (Calejo, 2004; Rho, 2004; Tarau, 2004). In Java Prolog Engine connectors libraries, the abstract engine is able to answer queries using the abstract term representation before mentioned. There are several implementation engines and in this project we try connect from top level engine to more concrete or specific Prolog Engine. Based on JPC we have a top level engine that communicate with more concretes engines. Over this concretes engines we offer several services to interact with the concrete engines with low coupling and platform independency. The common methods present in Prolog Engine are listed in the table 15.

| Return Type | Method and Description |
|---|---|
| `void` | **`abolish`**(**`String`** `functor,` `int arity`)<br>Remove all predicates that match with the predicate indicator (PI) formed by the concatenation of the given string functor and integer arity separated by slash (functor/arity). |
| `void` | **`asserta`**(**`PrologTerm`** `head,` **`PrologTerm...`** `body`)<br>Add a rule specified by the rule head and rule body if the specified rule clause non exist. |
| `void` | **`asserta`**(**`String`** `stringClause`)<br>Parse the string creating internal prolog clause and add the clause in the main memory program if the clause non exist. |
| `void` | **`assertz`**(**`PrologTerm`** `head,` **`PrologTerm...`** `body`)<br>Add a rule specified by the rule head and rule body if the specified rule clause non exist. |
| `void` | **`assertz`**(**`String`** `stringClause`)<br>Parse the string creating internal prolog clause and add the clause in the main memory program if the clause non exist. |
| `boolean` | **`clause`**(**`PrologTerm`** `head,` **`PrologTerm...`** `body`)<br>Find a rule specified by the rule head and rule body in main memory program that unify with the given clause returning true in this case.If the clause not exist in main memory program |

| | or exist but not unify with the given clause false value is returned. |
|---|---|
| boolean | **clause**(**String** stringClause)<br>Parse the string creating internal prolog clause and returning true if the clause in the main memory program unify with the given clause. |
| void | **consult**(**Reader** reader)<br>Consult a prolog program from specified reader parsing the prolog program and put this program into prolog engine. |
| void | **consult**(**String** path)<br>Consult a file specified by the string path loading an parsing the prolog program. |
| boolean | **contains**(**PrologTerm** goal, **PrologTerm...** goals)<br>Check if the given goal array have solution using the resolution engine mechanism. |
| boolean | **contains**(**String** goal)<br>Parse the string creating internal prolog clause and returning true if the given goal have solution using the resolution engine mechanism. |
| boolean | **currentOperator**(int priority, **String** specifier, **String** operator)<br>Check if in the wrapped prolog engine is defined some particular operator specified by your Priority, Specifier and Operator. |
| **Set**<**PrologOperator**> | **currentOperators**()<br>Operator set defined in the wrapped prolog engine. |
| boolean | **currentPredicate**(**String** functor, int arity)<br>Check if in the wrapped prolog engine is defined some particular predicate specified by your predicate indicator (PI = functor/arity). |
| **Set**<**PrologIndicator**> | **currentPredicates**()<br>Predicate set defined in the wrapped prolog engine. |
| void | **dispose**()<br>Clear program in main memory. |
| **Set**<**PrologIndicator**> | **getBuiltIns**()<br>Predicate set defined by the supported built-ins predicate in the wrapped prolog engine. |
| **String** | **getLicense**()<br>License of the wrapped engine. |
| **PrologLogger** | **getLogger**() |

| | | |
|---|---|---|
| | | Get the prolog system logger instance to report any errors or exceptions |
| `String` | `getName`() | |
| | Name of the wrapped engine. | |
| `String` | `getOSArch`() | |
| | Return the host operating system architecture. | |
| `String` | `getOSName`() | |
| | Return the host operating system name. | |
| `Set`<`PrologIndicator`> | `getPredicates`() | |
| | User defined predicate set defined in the wrapped prolog engine. | |
| `Set`<`PrologClause`> | `getProgramClauses`() | |
| | Make and return a copy of the clause set present in the current engine. | |
| `Map`<`String`,`List`<`PrologClause`>> | `getProgramMap`() | |
| | Make and return a copy of the clause map present in the current engine. | |
| int | `getProgramSize`() | |
| | Number of clauses in the current engine. | |
| `PrologProvider` | `getProvider`() | |
| | Get a Prolog provider instance hold in the current engine. | |
| `String` | `getVersion`() | |
| | Version of the wrapped engine. | |
| void | `include`(`Reader` reader) | |
| | Consult a prolog program from specified reader parsing the prolog program and include this program into current prolog engine. | |
| void | `include`(`String` path) | |
| | Consult a file specified by the string path loading an parsing the prolog program and include the loaded program into current engine. | |
| boolean | `isProgramEmpty`() | |
| | Check if the program in main memory is empty returning true if the clause number in the program is 0 and false in otherwise. | |
| `PrologClauseBuilder` | `newClauseBuilder`() | |
| | Create a new clause builder instance to build prolog clauses programmatically. | |
| `PrologQueryBuilder` | `newQueryBuilder`() | |
| | Create a new query builder instance to build prolog goal programmatically. | |
| void | `operator`(int priority, `String` specifier, `String` operator) | |

|  | Define an operator in the wrapped prolog engine with priority between 0 and 1200 and associativity determined by specifier according to the table below Specification table |
|---|---|
| void | **persist**(**String** path)<br>Save the prolog program present in the current engine to some specific file specified by string path. |
| void | **persist**(**Writer** writer)<br>Write the prolog clauses in program present in the current engine using the given writer. |
| **PrologQuery** | **query**(**PrologTerm**[] terms)<br>Create a new query being the goal the given prolog term array. |
| **PrologQuery** | **query**(**PrologTerm** term, **PrologTerm...** terms)<br>Create a new query with at least one prolog term goal. |
| **PrologQuery** | **query**(**String** query)<br>Create a new query being the goal the given string with prolog syntax. |
| **List**<**Map**<**String**,**PrologTerm**>> | **queryAll**(**PrologTerm** term, **PrologTerm...** terms)<br>Create a new prolog query and return the list of prolog terms that conform the solution set for the current query. |
| **List**<**Map**<**String**,**PrologTerm**>> | **queryAll**(**String** goal)<br>Create a new prolog query and return the list of prolog terms that conform the solution set for the current query. |
| **List**<**Map**<**String**,**PrologTerm**>> | **queryN**(int n, **PrologTerm** term, **PrologTerm...** terms)<br>Create a new prolog query and return the list of (N) prolog terms that conform the solution set for the current query. |
| **List**<**Map**<**String**,**PrologTerm**>> | **queryN**(int n, **String** goal)<br>Create a new prolog query and return the list of (N) prolog terms that conform the solution set for the current query. |
| **Map**<**String**,**PrologTerm**> | **queryOne**(**PrologTerm** term, **PrologTerm...** terms)<br>Create a new prolog query and return the prolog terms that conform the solution set for the current query. |
| **Map**<**String**,**PrologTerm**> | **queryOne**(**String** goal) |

| | Create a new prolog query and return the prolog terms that conform the solution set for the current query. |
|---|---|
| `void` | **`retract`**(**`PrologTerm`** `head,` **`PrologTerm...`** `body)`<br>Remove a rule specified by the rule head and rule body if the specified rule clause exist. |
| `void` | **`retract`**(**`String`** `stringClause)`<br>Parse the string creating internal prolog clause and remove the clause in the main memory program if the clause exist. |
| `boolean` | **`runOnLinux`**()<br>Check if the host operating system name refer to Linux OS. |
| `boolean` | **`runOnOSX`**()<br>Check if the host operating system name refer to OsX. |
| `boolean` | **`runOnWindows`**()<br>Check if the host operating system name refer to Windows OS. |
| `boolean` | **`unify`**(**`PrologTerm`** `t1,` **`PrologTerm`** `t2)`<br>Check that two terms (t1 and t2) unify. |

Tab. 15. Prolog Engine Interface Methods.

## Prolog Query

Prolog query is the mechanism to query the prolog database loaded in prolog engine. The way to create a new prolog query is invoking `query` () method in the Prolog Engine. When this method is called the prolog query is open an only `dispose` () in PrologQuery object close the current query and release all internal resources. Prolog query have several methods to manipulate the result objects. The main difference is in return types and result quantities. The result types enough depending of desire data type. Maps of variables name key and Prolog terms as value, Maps of variables name key and Java objects as value, List of before mentioned maps, Prolog terms array, Prolog terms matrix, list of Java Objects and list of list of Java Objects. Respect to result quantities Prolog query offer one, n-th or all possible solutions. This is an important feature because the Prolog engine is forced to retrieve the necessary solution quantities. Prolog query implement Iterable and Iterator. This implementation helps to obtain successive solutions present in the query. The common methods present in Prolog Query interface are listed in the table 8.

| Return Type | Method and Description |
|---|---|
| **`List`**<**`Map`**<**`String,PrologTerm`**>> | **`all`**()<br>Return a list of map of variables name key and Prolog terms as value that conform the solution set for the current query. |
| `List<List<Object>>` | **`allResults`**`()`<br>Return a list of list of Java Objects that conform the solution set for the current query. |
| `PrologTerm[][]` | **`allSolutions`**`()`<br>Return a Prolog terms matrix that conform the solution set for the current query. |
| `List<Map<String,Object>>` | **`allVariablesResults`**`()`<br>Return a list of map of variables name key and Java objects as value that conform the solution set for the current query. |
| `Map<String,PrologTerm>[]` | **`allVariablesSolutions`**`()`<br>Return an array of map of variables name key and Prolog terms as value that conform the solution set for the current query. |
| void | **`dispose`**`()`<br>Release all allocations for the query. |
| `PrologEngine` | **`getEngine`**`()`<br>Engine held by the current query. |
| `PrologProvider` | **`getProvider`**`()`<br>Provider instance |
| boolean | **`hasMoreSolutions`**`()`<br>Check if the current query has more solutions. |
| boolean | **`hasSolution`**`()`<br>Check that the current query has solution. |
| `PrologTerm[]` | **`nextSolution`**`()`<br>Return the next prolog terms solution array for the current query. |
| `Map<String,PrologTerm>` | **`nextVariablesSolution`**`()`<br>Return the next prolog terms that conform the solution set for the current query. |
| `PrologTerm[][]` | **`nSolutions`**`(int n)`<br>Return a Prolog terms matrix of n x m order that conform the solution set for the current query where n is the solution number and m is a free variable number in the query. |
| `Map<String,PrologTerm>[]` | **`nVariablesSolutions`**`(int n)`<br>Return an array of n size with maps of variables name key and Prolog terms as value that conform the solution set for the current query where n is the solution number. |
| `Map<String,PrologTerm>` | **`one`**`()`<br>Return a map of variables name key and Prolog terms as value that conform the solution set for the current query. |
| `List<Object>` | **`oneResult`**`()`<br>Return a list of Java objects that conform the solution set for the current query. |
| `PrologTerm[]` | **`oneSolution`**`()` |

|  | Return the prolog terms that conform the solution set for the current query. |
|---|---|
| `Map<String,Object>` | `oneVariablesResult()`<br>Return a map of variables name key and Java objects as value that conform the solution set for the current query. |
| `Map<String,PrologTerm>` | `oneVariablesSolution()`<br>Return the prolog terms that conform the solution set for the current query. |

Tab. 16. Prolog Engine Interface Methods.

```java
public class Main {
	public static void main(String[] args) {
		PrologProvider provider = Prolog.getProvider(SwiProlog.class);
		PrologEngine engine = provider.newEngine("zoo.pl");
		PrologVariable x = provider.newVariable("X", 0);
		PrologQuery query = engine.query(provider.newStructure("dark", x));
		while (query.hasNext()) {
			PrologTerm value = query.nextVariablesSolution().get("X");
			System.out.println(value);
		}
		query.dispose();
		engine.dispose();
	}
}

public class Main {
	public static void main(String[] args) {
		PrologProvider provider = Prolog.getProvider(SwiProlog.class);
		PrologEngine engine = provider.newEngine("zoo.pl");
		PrologVariable x = provider.newVariable("X", 0);
		PrologQuery query = engine.query(provider.newStructure("dark", x));
		for (Collection<PrologTerm> col : query) {
			for (PrologTerm prologTerm : col) {
				System.out.println(prologTerm);
			}
		}
		query.dispose();
		engine.dispose();
	}
}

public class Main {
	public static void main(String[] args) {
		PrologProvider provider = Prolog.getProvider(SwiProlog.class);
		PrologEngine engine = provider.newEngine("zoo.pl");
		PrologVariable x = provider.newVariable("X", 0);
		PrologQuery query = engine.query(provider.newStructure("dark", x));
		List<Object> solution = query.oneResult();
		for (int i = 0; i < solution.size(); i++) {
			System.out.println(solution.get(i));
		}
		query.dispose();
		engine.dispose();
	}
}
```

# Prolog Query Builder

Prolog query builder to create prolog queries. The mechanism to create a new query builder is using `PrologEngine.newQueryBuilder()`. The query builder emulates the query creation process. After define all participant terms with the `begin(PrologTerm)` method, we specify the first term in the query. If the query has more terms, they are created using `comma(PrologTerm)` for everyone. Clause builder have a `getQueryString()` for string representation of the clause in progress. After clause definition this builder have `query()` method that create the final query instance ready to be used. The follow code show how create a Prolog query `?- big(X), dark(X).` using PrologQueryBuilder interface. The table 17 show the prolog query builder interface methods.

| Return Type | Method and Description |
|---|---|
| **PrologQueryBuilder** | **begin**(**PrologTerm** term)<br>Append to the query builder the first term to be query. |
| **PrologQueryBuilder** | **begin**(**String** functor, **PrologTerm...** arguments)<br>Append to the query builder the first term to be query. |
| **PrologQueryBuilder** | **comma**(**PrologTerm** term)<br>Append to the query builder other term to be query in conjunctive mode. |
| **PrologQueryBuilder** | **comma**(**PrologTerm** left, **String** operator, **PrologTerm** right)<br>Append to the query builder other term to be query in conjunctive mode. |
| **PrologQueryBuilder** | **comma**(**String** functor, **PrologTerm...** arguments)<br>Append to the query builder other term to be query in conjunctive mode. |
| **PrologEngine** | **getEngine**()<br>Engine hold by the current builder |
| **String** | **getQueryString**()<br>Get the query in string format. |
| **PrologQuery** | **query**()<br>Create and return the result query. |
| **PrologQueryBuilder** | **semicolon**(**PrologTerm** term)<br>Append to the query builder other term to be query in disjunctive mode. |
| **PrologQueryBuilder** | **semicolon**(**PrologTerm** left, **String** operator, **PrologTerm** right)<br>Append to the query builder other term to be query in disjunctive mode. |
| **PrologQueryBuilder** | **semicolon**(**String** functor, **PrologTerm...** arguments) |

| | Append to the query builder other term to be query in disjunctive mode. |

Tab. 17. Prolog Engine Interface Methods.

```
PrologVariable x = provider.newVariable("X", 0);
PrologStructure big = provider.newStructure("big", x);
PrologStructure dark = provider.newStructure("dark", x);
PrologQueryBuilder builder = engine.newQueryBuilder();
PrologQuery query = builder.begin(dark).comma(big).query();
```

# Prolog Clause

Prolog clause is composed by two prolog terms that define a prolog clause, the head and the body. This representation considers the prolog clause body like a single term. If the body is a conjunctive set of terms, the body is a structure with functor/arity (, /2) and the first argument is the first element in the conjunction and the rest is a recursive functor/arity (, /2). The functor and arity for the clause is given from head term functor and arity. This class define some properties for commons prolog clause implementations. They are boolean flags that indicate if the prolog clause is dynamic multi-file and discontiguos. This class have several methods to access to the clause components and retrieve some clause properties and information about it. Additionally, this class contains a prolog provider reference for build terms in some operations. The table 18 show the methods present in Prolog clause interface.

| Return Type | Method and Description |
|---|---|
| **PrologTerm** | **getArgument**(int index)<br>Term located at some given index position in the clause head arguments. |
| **PrologTerm**[] | **getArguments**()<br>Term arguments present in the clause head. |
| int | **getArity**()<br>Integer number that represent the arguments number in the clause head. |
| **PrologTerm** | **getBody**()<br>Prolog term that represent the clause body. |
| **PrologTerm**[] | **getBodyArray**()<br>Get the clause body as terms array. |
| **Iterator**<**PrologTerm**> | **getBodyIterator**()<br>Iterator to iterate over all body terms. |
| **String** | **getFunctor**()<br>String that represent the functor in the clause head. |
| **PrologTerm** | **getHead**()<br>Prolog term that represent the clause head. |
| **String** | **getIndicator**()<br>Clause family functor/arity based indicator. |
| **PrologIndicator** | **getPrologIndicator**()<br>Clause family PrologIndicator based indicator. |
| **PrologTerm** | **getTerm**() |

| | Prolog term representation of the current clause. |
|---|---|
| boolean | **hasIndicator(String** functor, int arity) <br> Check if the current clause have functor/arity based indicator specified by arguments, false in otherwise. |
| boolean | **isDirective()** <br> True if this clause is a directive, false in other case. |
| boolean | **isDiscontiguous()** <br> **Deprecated.** <br> Natives engine don't offer information about that. |
| boolean | **isDynamic()** <br> **Deprecated.** <br> Natives engine don't offer information about that. |
| boolean | **isFact()** <br> True if this clause is a fact, false in other case. |
| boolean | **isMultifile()** <br> **Deprecated.** <br> Natives engine don't offer information about that. |
| boolean | **isRule()** <br> True if this clause is a rule, false in other case. |
| boolean | **unify(PrologClause** clause) <br> Check that two clauses unify. |

Tab. 18. Prolog Clause Interface Methods.

## Prolog Clause Builder

Prolog clause builder to create prolog clauses. The mechanism to create a new clause builder is using `PrologEngine.newClauseBuilder()`. The clause builder emulates the clause creation process. After define all participant terms with the `begin(PrologTerm)` method, we specify the head of the clause. If the clause is a rule, after head definition, the clause body is created with `neck(PrologTerm)` for the first term in the clause body. If the clause body have more terms, they are created using `comma(PrologTerm)` for everyone. Clause builder have a `getClauseString()` for string representation of the clause in progress. After clause definition this builder have `asserta()`, `assertz()`, `clause()`, `retract()` that use the wrapped engine invoking the correspondent methods for check, insert or remove clause respectively.

```
PrologTerm z = provider.newVariable("Z", 0);
PrologTerm darkZ = provider.newStructure("dark", z);
PrologTerm blackZ = provider.newStructure("black", z);
PrologTerm brownZ = provider.newStructure("brown", z);
PrologClauseBuilder builder = engine.newClauseBuilder();
```

```
builder.begin(darkZ).neck(blackZ).assertz();
builder.begin(darkZ).neck(brownZ).assertz();
```

The Prolog result in database is showed in the follow code. The table 19 show the Prolog clause builder interface methods.

```
dark(Z): -
      black(Z).
 dark(Z): -
      brown(Z).
```

| Return Type | Method and Description |
|---|---|
| void | **asserta**()<br>Add the current clause in the main memory program if the clause non exist. |
| void | **assertz**()<br>Add the clause in the main memory program if the clause non exist. |
| **PrologClauseBuilder** | **begin**(**PrologTerm** term)<br>Append to the clause builder the head term in the clause. |
| **PrologClauseBuilder** | **begin**(**String** functor, **PrologTerm**... arguments)<br>Append to the clause builder the head term in the clause. |
| boolean | **clause**()<br>Check if the clause in the main memory program unify with the current clause and return true. |
| **PrologClauseBuilder** | **comma**(**PrologTerm** term)<br>Append to the clause builder other term in the clause body in conjunctive mode. |
| **PrologClauseBuilder** | **comma**(**PrologTerm** left, **String** operator, **PrologTerm** right)<br>Append to the clause builder other term in the clause body in conjunctive mode. |
| **PrologClauseBuilder** | **comma**(**String** functor, **PrologTerm**... arguments)<br>Append to the clause builder other term in the clause body in conjunctive mode. |
| **String** | **getClauseString**()<br>Get the clause in string format. |
| **PrologEngine** | **getEngine**()<br>Engine hold by the current builder |
| **PrologClauseBuilder** | **neck**(**PrologTerm** term)<br>Append to the clause builder the first term in the clause body. |

| `PrologClauseBuilder` | `neck(PrologTerm left,`<br>`String operator,`<br>`PrologTerm right)`<br>Append to the clause builder the first term in the clause body. |
|---|---|
| `PrologClauseBuilder` | `neck(String functor,`<br>`PrologTerm... arguments)`<br>Append to the clause builder the first term in the clause body. |
| `void` | `retract()`<br>Remove the clause in the main memory program if the clause exist. |

Tab. 19. Prolog Clause Interface Methods.

# Prolog Logger

Prolog Logger is the logger platform interface to log message at any level. Is an adapter for Logger adapting the Java logger mechanism for use with the most popular logger methods. This logger mechanism is accessible from `PrologProvider.getLogger()` or `PrologEngine.getLogger()` This logger interface have all traditional methods used to log messages at different levels (trace, debug, info,warn,error). The levels used for this logger interface are Level constants present in the table.

| Method | Level |
|---|---|
| `trace(Object, Object, Throwable)` | `Level.FINEST` |
| `debug(Object, Object, Throwable)` | `Level.FINE` |
| `info(Object, Object, Throwable)` | `Level.INFO` |
| `warn(Object, Object, Throwable)` | `Level.WARNING` |
| `error(Object, Object, Throwable)` | `Level.SEVERE` |

Tab. 20. Prolog Logger methods and levels.

By default, the platform implements a logger mechanism for drop log messages in Operating System temporal directory into files named prolobjectlink-YYYY.MM.DD. In AbstractLogger class there are many implementations for this interface. Every final implementation class can extend from AbstractLogger.

# Prolog Scripting in Java

Java 6 added scripting support to the Java platform that lets a Java application execute scripts written in scripting languages such as Rhino JavaScript, Groovy, Jython, JRuby, Nashorn JavaScript, etc. All classes and interfaces in the Java Scripting API are in the javax.script package. Using a scripting language in a Java application provides several advantages, dynamic type, simple way to write programs, user customization, easy way to develop and provide domain-specific features that are not available in Java. For achieve this propose Java Scripting API introduce a scripting engine component. A script engine is a software component that executes programs written in a particular scripting language. Typically, but not necessarily, a script engine is an implementation of an interpreter for a scripting language. To run a script in Java is necessary perform the following three steps, create a script engine manager, get an instance of a script engine from the script engine manager and Call the `eval()` method of the script engine to execute a script.

```java
public class Main {
    public static void main(String[] args) {
        ScriptEngineManager manager = new ScriptEngineManager();
        ScriptEngine engine = manager.getEngineByName("prolog");
        Boolean result = engine.eval("?- X is 5+3.");
        Integer solution = engine.get("X");
        System.out.println(solution);
    }
}
```

Using script engine, it possible read Prolog source file. Read Prolog source file allow coding all prolog source in separate mode respect to Java program.

```java
public class Main {
    public static void main(String[] args) {
        ScriptEngineManager manager = new ScriptEngineManager();
        ScriptEngine engine = manager.getEngineByName("prolog");
        Boolean read = engine.eval(new FileReader("family.pl"));
        Boolean eval = engine.eval("?- parent( Parent, Child)");
        Object parent = engine.get("Parent");
        Object child = engine.get("Child");
        System.out.println(parent);
        System.out.println(child);
    }
}
```

## Analysis and Result

JPC and JPI Prolog Engine share many services that are common for all engines implementations. This services are grouped by two categories, Consults methods and Modifiers methods. Consult methods using the engine to query for anything, a file, clauses, and compute solutions. Modifiers methods using the engine for modify a file and clauses. The table 21 show a comparison between JPC and JPI Prolog Engine services.

| Description | JPC | JPI |
|---|---|---|
| Clear program in main memory and release engine resources. | close | dispose |
| Check if the goal have solution using the resolution engine mechanism or execute a Prolog command. | command | contains |
| Create a new query with at least one prolog term goal. | query | query |
| Convert a Java Object to term | toTerm | toTerm |
| Convert a term to Java Object | fromTerm | fromTerm |
| Define a prolog flag | setPrologFlag | - |
| Check the prolog flag definition. | currentPrologFlag | - |
| Define a prolog operator | currentOp | operator |
| Add a clause in the prolog program at first in the clause family | asserta | asserta |
| Add a clause in the prolog program at final in the clause family | assertz | assertz |
| Remove a clause in the prolog program | retract | retract |
| Remove all clause | retractAll | - |
| Remove all clause that match with the predicate indicator (functor/arity). | abolish | abolish |
| Check the clause existence in prolog program. | clause | clause |
| Consult a file | ensured_loaded | consult |
| Finds each solution to some goal considering the free variables | bagof | - |
| Finds each solution to some goal | findall | queryAll |
| Finds each solution to some goal and remove duplicate and sort the solution set | setof | - |
| | forall | - |
| Check that two terms (x and y) unify. | unify | unify |
| Check the prolog operator definition. | isOperator | currentOperator |
| Save the prolog program present in the current engine to some specific file | - | persist |

Tab. 21. Comparison between JPC and JPI Prolog Abstract Engines.

JPI project have implemented a benchmark to compare the throughput of the different Prolog Engines using JPI interface. This Prolog benchmark is based on CIAO Prolog (Hermenegildo, y otros, 2012) distribution benchmark. Every benchmark program evaluates a Prolog engine feature in time unit. The follow table explain each Prolog benchmark program.

| Benchmark | Description |
|---|---|
| boresea | Show the effect of procedure pure calls. Can be called the peak performance of the prolog system. |
| choice_point | Tests call invoking the creation of a choice point, i.e. a branch point where the execution will possibly come back to in case of backtracking. |
| backtrack1 | Exhibits a kind of backtracking called " deep" |
| backtrack2 | Exhibits a kind of backtracking called " shallow" |
| cut_100_times | Contains a lot of cut at execution time. |

| dereference | Program to benchmark the dereferencing speed. |
|---|---|
| enviroment | Attempts to evaluate the creation and deletion of environments. |
| index_clause | Test for clause indexing that is the selection of a clause due to the type of an argument. |
| create_list | Programs to evaluate the unification process in the Prolog system. |
| create_struct | |
| match_list | |
| match_struct | |
| unification | |

Tab. 22. Prolog Benchmark programs and description.

The Java Prolog Benchmark (JPB) is an implementation of before mentioned benchmark over JPI. JPB use the JMH library in version 1.19 over Eclipse OpenJ9 VM-1.8.0_192 on Windows 10 Home x64. JMH is a Java harness for building, running, and analyzing Nano, Micro and Macro benchmarks written in Java and other languages targeting the JVM. JPB was executed on Lenovo Ideapad 110 laptop with a ADM A6-7310 with AMD Radeon R4 Graphics processor to 2.0 GHz, 4 GB of RAM memory and 120 GB SSD. The table 23 show the different result for every Prolog Engine and benchmark program separate by a dot. This table show the minimum, average, maximum, error and standard deviation. The unit for the result values are milliseconds by operations (ms/op).

| Benchmark | Min | Ave | Max | Error | Stdev |
|---|---|---|---|---|---|
| JTrolog.dereference | 4.658 | 5.169 | 7.294 | 0.590 | 0.679 |
| JLog.dereference | 26.523 | 29.010 | 33.049 | 1.624 | 1.870 |
| Swi7.dereference | 82.208 | 118.523 | 270.427 | 44.372 | 51.098 |
| Swi.dereference | 119.436 | 124.691 | 134.840 | 3.426 | 3.945 |
| TuProlog.dereference | 129.510 | 160.637 | 187.452 | 15.469 | 17.814 |
| | | | | | |
| JTrolog.createList | 0.361 | 0.434 | 0.531 | 0.043 | 0.049 |
| TuProlog.createList | 0.892 | 2.057 | 4.527 | 1.074 | 1.237 |
| JLog.createList | 25.297 | 46.589 | 135.373 | 24.452 | 28.159 |
| Swi7.createList | 84.783 | 113.168 | 181.969 | 23.236 | 26.758 |
| Swi.createList | 121.373 | 6637.999 | 129588.099 | 25129.795 | 28939.509 |
| | | | | | |
| JTrolog.choicePoint | 0.232 | 0.318 | 0.533 | 0.059 | 0.068 |
| TuProlog.choicePoint | 0.856 | 1.625 | 5.038 | 0.785 | 0.904 |
| JLog.choicePoint | 27.529 | 43.303 | 168.283 | 26.484 | 30.499 |
| Swi7.choicePoint | 80.870 | 83.414 | 93.411 | 2.739 | 3.155 |
| Swi.choicePoint | 126.580 | 258.866 | 302.271 | 41.326 | 47.591 |
| | | | | | |
| JTrolog.enviroment | 0.275 | 0.318 | 0.404 | 0.031 | 0.036 |
| TuProlog.enviroment | 0.916 | 1.334 | 2.562 | 0.278 | 0.320 |
| JLog.enviroment | 25.786 | 42.578 | 149.087 | 24.232 | 27.905 |
| Swi7.enviroment | 84.218 | 103.194 | 132.318 | 14.495 | 16.693 |
| Swi.enviroment | 119.406 | 123.578 | 128.415 | 2.278 | 2.624 |
| | | | | | |
| JTrolog.unification | 0.301 | 0.400 | 1.015 | 0.148 | 0.171 |
| TuProlog.unification | 0.547 | 0.654 | 0.978 | 0.116 | 0.134 |
| JLog.unification | 27.203 | 40.454 | 93.933 | 16.783 | 19.327 |
| Swi7.unification | 82.544 | 96.754 | 189.181 | 20.738 | 23.881 |
| Swi.unification | 124.718 | 402352.360 | 8041075.315 | 1561276.969 | 1797968.879 |
| | | | | | |
| JTrolog.backtrack1 | 0.235 | 0.379 | 1.015 | 0.171 | 0.197 |
| TuProlog.backtrack1 | 0.523 | 0.558 | 0.641 | 0.030 | 0.034 |
| JLog.backtrack1 | 30.867 | 46.213 | 175.463 | 28.616 | 32.954 |
| Swi7.backtrack1 | 80.646 | 86.140 | 114.160 | 6.675 | 7.687 |
| Swi.backtrack1 | 122.080 | 129.405 | 146.025 | 4.273 | 4.921 |
| | | | | | |
| JTrolog.boresea | 0.388 | 0.413 | 0.475 | 0.020 | 0.023 |
| TuProlog.boresea | 0.730 | 1.103 | 1.820 | 0.367 | 0.423 |
| JLog.boresea | 25.014 | 41.428 | 105.256 | 14.895 | 17.153 |
| Swi7.boresea | 80.504 | 82.625 | 100.992 | 3.834 | 4.415 |
| Swi.boresea | 124.135 | 143.347 | 290.281 | 38.662 | 44.523 |
| | | | | | |
| JTrolog.backtrack2 | 0.263 | 0.280 | 0.302 | 0.010 | 0.012 |
| TuProlog.backtrack2 | 0.332 | 0.371 | 0.715 | 0.072 | 0.082 |
| JLog.backtrack2 | 27.425 | 30.031 | 33.087 | 1.352 | 1.557 |
| Swi7.backtrack2 | 80.711 | 82.688 | 99.415 | 3.515 | 4.048 |
| Swi.backtrack2 | 123.665 | 145.707 | 272.748 | 37.636 | 43.341 |
| | | | | | |
| JTrolog.benchQueryAll | 0.499 | 0.601 | 0.730 | 0.056 | 0.065 |
| TuProlog.benchQueryAll | 2.608 | 3.030 | 4.597 | 0.500 | 0.575 |
| JLog.benchQueryAll | 28.311 | 36.254 | 44.927 | 4.840 | 5.574 |
| Swi7.benchQueryAll | 81.132 | 95.479 | 243.213 | 31.351 | 36.104 |
| Swi.benchQueryAll | 123.917 | 130.416 | 153.614 | 5.545 | 6.386 |
| | | | | | |
| JTrolog.indexClause | 0.248 | 0.293 | 0.684 | 0.081 | 0.093 |
| TuProlog.indexClause | 1.469 | 1.771 | 3.948 | 0.578 | 0.665 |
| JLog.indexClause | 29.863 | 41.806 | 97.886 | 12.937 | 14.899 |
| Swi7.indexClause | 85.409 | 118.210 | 359.058 | 55.061 | 63.408 |
| Swi.indexClause | 130.373 | 143.717 | 214.025 | 14.608 | 16.823 |
| | | | | | |
| JTrolog.benchQuery | 0.254 | 0.291 | 0.399 | 0.037 | 0.043 |
| TuProlog.benchQuery | 1.145 | 1.385 | 3.299 | 0.446 | 0.513 |
| JLog.benchQuery | 30.035 | 44.767 | 94.387 | 12.163 | 14.007 |
| Swi7.benchQuery | 81.336 | 83.789 | 96.867 | 3.140 | 3.616 |
| Swi.benchQuery | 128.774 | 150.243 | 276.261 | 41.693 | 48.014 |
| | | | | | |
| JTrolog.cut100Times | 0.161 | 0.276 | 1.176 | 0.187 | 0.216 |
| TuProlog.cut100Times | 0.558 | 0.630 | 0.907 | 0.077 | 0.089 |
| JLog.cut100Times | 27.196 | 32.433 | 58.083 | 7.054 | 8.124 |
| Swi7.cut100Times | 84.459 | 121.986 | 416.699 | 67.764 | 78.037 |
| Swi.cut100Times | 119.974 | 140.565 | 271.021 | 37.051 | 42.668 |
| | | | | | |
| JTrolog.matchList | 0.450 | 0.534 | 0.707 | 0.047 | 0.054 |
| TuProlog.matchList | 2.803 | 3.733 | 5.648 | 0.744 | 0.857 |
| JLog.matchList | 27.196 | 44.256 | 107.744 | 17.797 | 20.495 |
| Swi7.matchList | 82.490 | 108.594 | 220.089 | 26.056 | 30.006 |
| Swi.matchList | 241.915 | 262.250 | 282.613 | 9.630 | 11.090 |
| | | | | | |
| JTrolog.matchStruct | 0.509 | 0.552 | 0.661 | 0.034 | 0.040 |
| TuProlog.matchStruct | 2.777 | 3.626 | 5.030 | 0.697 | 0.803 |
| JLog.matchStruct | 26.844 | 43.180 | 89.131 | 14.318 | 16.488 |
| Swi7.matchStruct | 94.334 | 105.181 | 159.194 | 15.319 | 17.641 |
| Swi.matchStruct | 240.859 | 257.967 | 266.426 | 5.057 | 5.824 |
| | | | | | |
| JTrolog.choicePoint0Arg | 0.187 | 0.234 | 0.700 | 0.097 | 0.112 |
| TuProlog.choicePoint0Arg | 1.064 | 1.321 | 3.544 | 0.464 | 0.535 |
| JLog.choicePoint0Arg | 27.418 | 40.847 | 111.714 | 15.917 | 18.330 |
| Swi7.choicePoint0Arg | 81.250 | 101.547 | 207.563 | 28.537 | 32.863 |
| Swi.choicePoint0Arg | 126.067 | 353.090 | 1850.143 | 316.775 | 364.799 |
| | | | | | |
| JTrolog.createStruct | 0.337 | 0.388 | 0.441 | 0.026 | 0.030 |
| TuProlog.createStruct | 1.385 | 2.121 | 3.104 | 0.584 | 0.672 |
| JLog.createStruct | 26.389 | 40.845 | 71.522 | 10.631 | 12.242 |
| Swi7.createStruct | 83.764 | 90.187 | 105.465 | 6.170 | 7.105 |
| Swi.createStruct | 119.428 | 139.177 | 440.288 | 61.581 | 70.917 |
| | | | | | |
| JTrolog.enviroment0Arg | 0.255 | 0.337 | 0.715 | 0.100 | 0.115 |
| TuProlog.enviroment0Arg | 0.595 | 0.842 | 1.161 | 0.102 | 0.117 |
| JLog.enviroment0Arg | 29.061 | 39.678 | 82.503 | 10.595 | 12.201 |
| Swi7.enviroment0Arg | 81.122 | 106.385 | 164.710 | 19.022 | 21.906 |
| Swi.enviroment0Arg | 133.933 | 141.338 | 145.693 | 2.825 | 3.253 |

Tab. 23. Comparison between JPI Prolog Benchmarks.

Java-based engine implementation has better throughput because use a direct memory access. Natives implementation like JPI-SWI and JPI-SWI7 use a cache file. The performance is affected by the time to access to cache file. The best Java-based engine implementation is JTrolog. JTrolog has the best times over every Prolog benchmark program. The second best time for every Prolog benchmark program is TuProlog except in dereference Prolog benchmark program. TuProlog have the worst time in dereferencing speed. JLog is the middle engine by throughput. The most remarkable feature is the dereferencing speed occupying the second place in this feature. The last one by throughput is SWI. SWI is the last in every Prolog benchmark program except in dereference Prolog benchmark program. The slowest Prolog benchmark program for SWI is unification. This test is the main cause of the SWI slow performance. SWI7 is a most efficient version of SWI-Prolog. Have a better performance respect to your older version because optimize the unification procedure. This optimization reduces the time for unification when JPI is used. The table 24 summarize the average measure for every engine implementation. By the total average the best Prolog engine is JTrolog. After JTrolog the best Prolog engines performance is for TuProlog, JLog, SWI7 and SWI by this order.

| JTrolog | JLog | Swi | Swi7 | TuProlog |
|---------|----------|----------|----------|----------|
| 11.21689 | 683.6709 | 411634.7 | 1697.866 | 186.7981 |

Tab. 24. Average total of JPI Prolog Benchmarks.

The figure 1 show the performance comparison chart - The Y-Axis represents normalized score in logarithmic scale - lower is better.

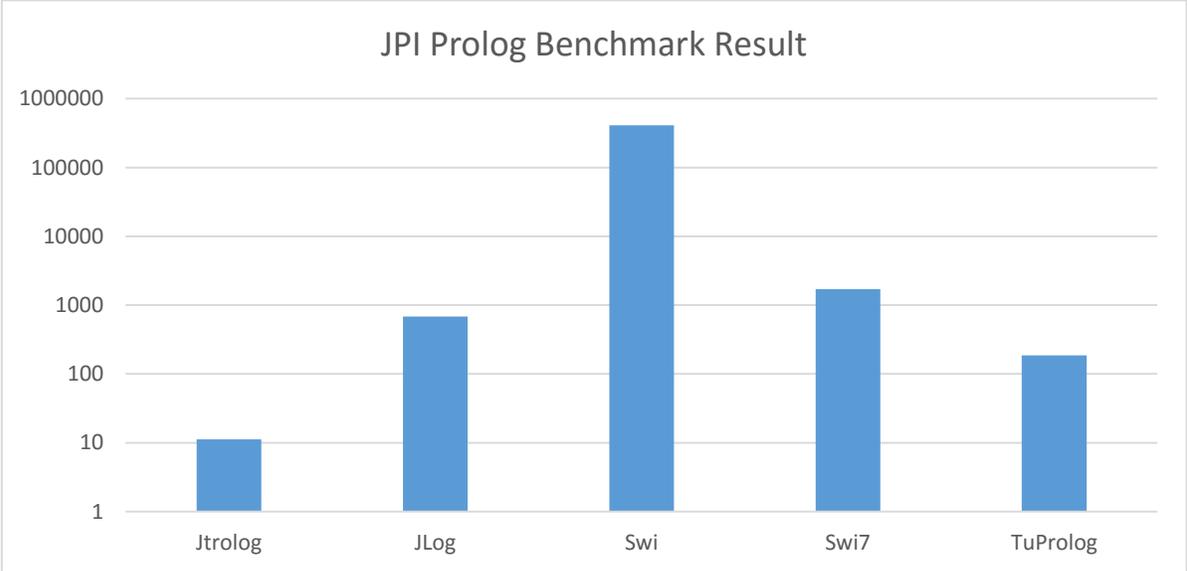

Fig. 1. Average total of JPI Prolog Benchmarks in logarithmic scale.

# Conclusions

In this paper we presented a simple Prolog interface written in Java, which connect Java and Prolog programming languages named Java Prolog Interface (JPI). JPI is an intent to connect all prolog implementations, native engines like SWI, YAP and XSB. and Java-Based engines like TuProlog, JLog and JTrolog. JPI s influenced by Interprolog, PDT and JPC. At the moment only SWI, TuProlog, JLog and JTrolog are the well implemented for production. Projects like JiProlog, YAP and XSB are in development mode. The project planning in the future the develop of more Prolog engine connectors like CIAO, ECLiPSe (Schimps & Shen, 2012), Jekejeke (XLOG, 2010) and others. The main goal is cover all Prolog engines that have a Java connector or is Java-based.

JPI provides clear and concise access to Prolog and simplifies the integration of predicates in Prolog and provides an automated object-to-term mapping mechanism. Allow interact with concrete engines in transparent way and single sourcing for every Prolog Provider. Allow the coupling and decoupling different concrete Prolog engines implementations and the application code still remain the same. Provide data type abstraction. Introduce Java Scripting Engine implementation. Prolog and Java connectors need more over Java Platform. Java Prolog Interface need be a specific API specification like (`javax.logic`) for all logic and functional language interaction. One specification like this allow more direct implementations reducing the impedance mismatch cost and self-platform distribution API.